\documentclass[pra,aps,superscriptaddress,twocolumn,color]{revtex4-2}

\usepackage{amsmath}
\usepackage{amsfonts}
\usepackage{bm}
\usepackage{graphicx}
\usepackage{array,color}
\usepackage{braket}
\usepackage[table]{xcolor}

\begin{document}
\title{Phase separation and multistability of two-component Bose-Einstein condensate in an optical cavity}

\author{Abid Ali}
\affiliation{Department of Engineering Science, University of
Electro-Communications, Tokyo 182-8585, Japan}

\author{Farhan Saif}
\affiliation{Department of Electronics, Quaid-i-Azam University, 45320, Islamabad, Pakistan}

\author{Hiroki Saito}
\affiliation{Department of Engineering Science, University of
Electro-Communications, Tokyo 182-8585, Japan}

\date{\today}

\begin{abstract}
We examine the multistability associated with miscibility-immiscibility conditions for a two-component Bose-Einstein condensate coupled to the light field in an optical cavity. For a strongly immiscible condition, the system exhibits a variety of density structures, including separated state, stripe state, and their coexistence. The multistability arises from these spatial structures of the two-component condensate, which significantly alter the hysteresis curve with respect to the intensity of cavity pumping. We present a variational approach to confirm our numerical results.

\end{abstract}

\maketitle

\section{Introduction}
\label{sec:first}
Phase separation is a well-known phenomenon in nature, and the immiscibility of oil and water is the most prominent example. The experimental realization of two-component Bose-Einstein condensates (BECs)  \cite{myatt1997production,hall1998dynamics,stamper1998optical,modugno2001bose} provided an ideal platform to investigate the miscible-immiscible transition in quantum systems. Later on, efforts were made to create two-component BECs with different atomic species, e.g., $^{41}$K-$^{87}$Rb \cite{modugno2001bose}, $^{7}$L-$^{133}$Cs \cite{mudrich2002sympathetic}, $^{87}$Rb-$^{84}$Sr and $^{87}$Rb-$^{88}$Sr \cite{pasquiou2013quantum}. In recent years, the interest in two-component BECs has been revived by more controlled and more precise experimental results demonstrating the phenomenon of phase separation \cite{mertes2007nonequilibrium,papp2008tunable,tojo2010controlling,eto2016nonequilibrium}.

Preparing BECs in optical cavities has opened new frontiers for exploring the light-matter interaction, where the cavity field couples predominantly to cold atoms.
Ultracold atoms play a crucial role inside the optical cavity, in which the atom-photon interaction induces an optical lattice for atoms, while the induced matter-wave grating changes the cavity resonance. In recent years, BECs loaded in optical cavities have been studied extensively \cite{brennecke2008cavity,mottl2012roton,schmidt2014dynamical,klinder2015observation,piazza2015self,landig2016quantum,leonard2017supersolid,kollar2017supermode,vaidya2018tunable,schuster2020supersolid,li2021first,guo2021optical}.
Atom-photon interactions in a cavity-BEC system generate highly nonlocal nonlinearity, which gives rise to many interesting phenomena, such as self-organization of atoms inside the optical cavity \cite{nagy2008self,guo2019emergent,ostermann2020unraveling}, Dicke quantum phase transitions \cite{baumann2010dicke,nagy2010dicke,baumann2011exploring,bhaseen2012dynamics,klinder2015dynamical}, optical bistability                                                                       
 \cite{black2003observation,gupta2007cavity,colombe2007strong,zhang2009nonlinear,szirmai2010quantum,yang2011controllable,dalafi2013controllability,dalafi2017intrinsic}, instability and chiral dynamics \cite{dogra2019dissipation}, 
and Floquet dynamics \cite{luo2018self,li2019nonlinear}.
 
In previous studies, the bistable behaviors of single-component BECs in optical cavities have been extensively studied both theoretically \cite{zhang2009nonlinear,szirmai2010quantum,yang2011controllable,dalafi2013controllability,dalafi2017intrinsic} and experimentally \cite{black2003observation,gupta2007cavity,colombe2007strong}. Two-component and spinor BECs in cavities have also been studied by many researchers \cite{zhou2009cavity,huang2010modulational,zhou2010spin,zheng2011cavity,zhou2011vortex,dong2011multistability,safaei2013bistable,mivehvar2017disorder,masson2017cavity,landini2018formation,chiacchio2019dissipation,davis2019photon,buvca2019dissipation}. 
However, most of these studies ignore the atom-atom interaction, which determines the miscibility of multicomponent BECs.
Since bistability is strongly affected by the density distribution of the BEC, we expect that the miscible-immiscible transition of a two-component BEC plays a vital role in bistability and multistability, which the present paper focuses on. We will show that the global structure of the condensate changes abruptly when the transition occurs between multistable branches. We find a variety of multistable phases, such as a separated phase, alternate stripe phase, and their coexistence. This multistability is sensitive to the inter-component interaction strength.
We will also analyze the multistability using a variational approach to confirm our numerical results.

This paper is organized as follows. In Sec.~\ref{sec:second}, we present the mean-field description of the system based on coupled Gross-Pitaevskii (GP) and cavity field equations. 
In Sec.~\ref{sec:third}, we present numerical results for the multistability in the absence of a trapping potential.  
In Sec.~\ref{sec:fourth}, we propose a variational method to understand the numerical results. In Sec.~\ref{sec:fifth}, we investigate a harmonically trapped system.
Finally, we summarize our results in Sec.~\ref{sec:sixth}.

\section{Model}
\label{sec:second}
\begin{figure}[tb]
\includegraphics[scale=0.5]{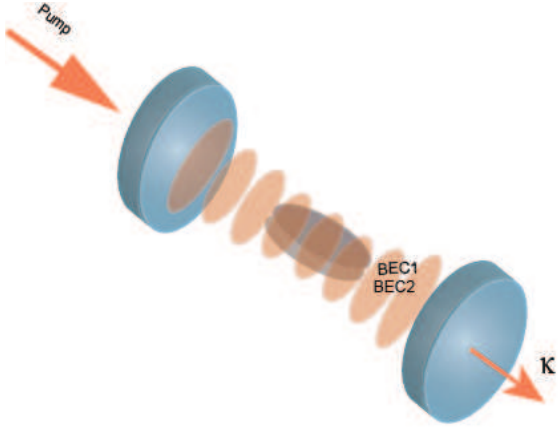} 
\caption{Schematic illustration of the two-component BEC-cavity system driven by laser pumping along the cavity axis. The cavity field is pumped at a rate of $\eta$ and decays at a rate of $\kappa$. The cavity mode near the BECs is approximated as $\cos(kx)$.}
\label{fig:model}
\end{figure}

We consider a two-component BEC located inside a high-Q single-mode optical cavity, as illustrated in Fig.~\ref{fig:model}. The cavity mode frequency is $ \omega_{c} $, and the transition frequency of the two-level atoms is $ \omega_{aj} $ with components $j = 1$ and 2. The coupled BEC-cavity system is coherently driven by the external laser pump along the cavity axis with frequency $ \omega_{p} $ and amplitude $\eta$, and the cavity field decays at a rate of $\kappa$. The atom-pump detuning and the cavity-pump detuning are represented as $ \Delta_{aj}=\omega_{aj}-\omega_{p} $ and $ \Delta_{c}=\omega_{c}-\omega_{p} $, respectively. We will assume that the laser pump is detuned far from the atomic transition $ \omega_{aj} $ to ensure that the electronic excitation is sufficiently low and the upper atomic excited state can be eliminated adiabatically.

In the mean-field description, the time evolution of macroscopic wavefunctions $ \Psi_{1}(\bm{r}, t) $ and $ \Psi_{2}(\bm{r}, t) $ of components $ 1 $ and $ 2 $ are described by the GP equation, including their interaction with the cavity field. We assume that the cavity mode at the location of the BECs can be approximated by the one-dimensional sinusoidal function $\cos(kx)$, where $k$ is the wave-number of the cavity mode. We also assume that the cavity field can be treated as a classical field $\alpha = \langle \hat a \rangle$, where $\hat a$ is the annihilation operator of the cavity photon. The coupled GP equations for the macroscopic wave functions are then given by
\begin{equation}
\begin{aligned}
i\hbar\frac{\partial\Psi_{1}}{\partial t} ={} & \Big[-\frac{\hbar^{2}}{2m_{1}}          \nabla^{2}+V_{1}(\bm{r})+U_{1}\vert\alpha\vert^{2} \cos^{2}(kx)  \\
                                              & +g_{11}\vert\Psi_{1}\vert^{2}+g_{12}\vert   \Psi_{2}\vert^{2} \Big]\Psi_{1},
\end{aligned}
\label{eq:one}
\end{equation}
\begin{equation}
\begin{aligned}
i\hbar\frac{\partial\Psi_{2}}{\partial t} ={} & \Big[-\frac{\hbar^{2}}{2m_{2}}          \nabla^{2}+V_{2}(\bm {r})+U_{2}\vert\alpha\vert^{2} \cos^{2}(kx)  \\
                                              & +g_{22}\vert\Psi_{2}\vert^{2}+g_{12}\vert   \Psi_{1}\vert^{2} \Big] \Psi_{2},
\end{aligned}
\label{eq:two}
\end{equation}
and the equation of motion for the cavity field has the form
\begin{equation}
\begin{aligned}
\frac{d \alpha}{d t} ={} & -i\Big[ \Delta_c+
U_{1}\int d\bm{r}\vert\Psi_{1}\vert^{2}   \cos^{2}(kx) \\
                                       & + U_{2}\int d\bm{r}\vert\Psi_{2}\vert^{2}\cos^{2}(kx) \Big]\alpha - \kappa\alpha+\eta,
\end{aligned}
\label{eq:three}
\end{equation}
where $g_{jj'}=\dfrac{2\pi \hbar^{2}a_{jj'}}{m_{jj'}}$ is the intercomponent interaction coefficient with $m_{jj'}=(m_j^{-1}+m_{j'}^{-1})^{-1}$ the reduced mass and $a_{jj'}$ being the $s$-wave scattering length between components $j$ and $j'$, $V_{j}$ is the external potential for component $j$, and $U_{j} = -g_{0j}^{2} / \Delta_{a j}$ is the maximal light shift per photon that an atom may experience, with $g_{0j}$ being the atom-photon coupling constant for component $j$. The wave function is normalized as $\int d\bm{r} |\Psi_j(\bm{r}, t)|^2 = N_j$, where $N_j$ is the number of atoms in component $j$.

In experiments, the cavity damping is much faster than the mechanical motion of the condensate. We therefore assume that the cavity field follows the condensate adiabatically. Setting $\frac{d\alpha}{dt} = 0 $ in Eq.~(\ref{eq:three}), the instantaneous photon number is obtained as
\begin{equation}
\vert {\alpha}{(\text{t})} \vert^{2}=
    \frac{\eta^{2}}{\Big[\delta_{c} + \sum_{j=1}^2 \frac{U_{j}}{2}\int d\bm{r}\vert\Psi_{j}(\bm{r},t)\vert^{2}   \cos(2kx)\Big]^{2}+\kappa^{2}}.
\label{eq:four}
\end{equation}
where $\delta_{c}=\Delta_{c}+\frac{N_{1}U_{1}}{2}+\frac{N_{2}U_{2}}{2}$. Substituting Eq.~(\ref{eq:four}) into Eqs.~(\ref{eq:one}) and ~(\ref{eq:two}), we obtain a nonlocal GP equation. The nonlinearity arises not only from the atom-atom interaction but also from the atom-photon interaction, which causes an effective nonlocal interaction between atoms, leading to the emergence of bistability. Bistability phenomena have already been studied in the system of a single-component BEC and optical cavity   \cite{zhang2009nonlinear,szirmai2010quantum,yang2011controllable,dalafi2013controllability,dalafi2017intrinsic,black2003observation,gupta2007cavity,colombe2007strong}. 

The miscibility of the two components is important in this system, which is determined by the interaction coefficients $g_{ij}$. In the following study, we assume that all the atom-atom interactions are repulsive, $g_{ij} > 0$. In this case, for a homogeneous system without a cavity field, the two components are immiscible and phase separation occurs, when $g_{11} g_{22} < g_{12}^2$ is satisfied \citep{becbook}. In the presence of the cavity field, the density distribution of the BEC is changed by the optical potential and this miscibility condition is not simply applicable.

We convert the coupled GP equations into their dimensionless forms, and all the quantities in the following are dimensionless. We normalize the length and time by $ k^{-1} $ and $ (\hbar k^{2}/ 2m)^{-1} $, respectively, where we assume $m_1 = m_2 \equiv m$ for simplicity. We also assume $g_{11} = g_{22} \equiv g$ to reduce the parameter space. The wave function $\Psi_j$ is scaled by $\sqrt{N k^{3}}$, and $\int d\bm{r} |\Psi_j|^2 = N_j / N$ is satisfied in the dimensionless unit, where $N = N_1 + N_2$ is the total number of atoms. In the following study, we assume $N_1 = N_2 = N / 2$.

\section{Numerical Simulation of an ideal system}
\label{sec:third} 
We consider an ideal system to focus on the multistability of the system, where the trapping potential is absent, i.e., $ V_{1} = V_{2} = 0 $, and the periodic boundary condition is imposed. For simplicity, we consider a two-dimensional system with size $(2\pi \ell)^2$ with $\ell = 4$, and the number of atoms per $(2\pi)^2$ area is defined as $\tilde{N} = N / \ell^2$. We also define $\tilde{g} = g \tilde{N}$ and $\tilde{g}_{12} = g_{12} \tilde{N}$. We solve the imaginary-time propagation of Eqs.~(\ref{eq:one}) and~(\ref{eq:two}) with Eq.~(\ref{eq:four}) numerically, using the pseudospectral method \cite{press2007numerical}, where $i$ on the left-hand sides of Eqs.~(\ref{eq:one}) and ~(\ref{eq:two}) is replaced with $-1$. In the numerical simulations, we discretize the space into a $256\times256$ mesh, and the time step is $dt=0.001$.  We add a small random number to each mesh of the initial state to break the symmetry of the system.

\begin{figure*}[tb]
\includegraphics[scale=0.5]{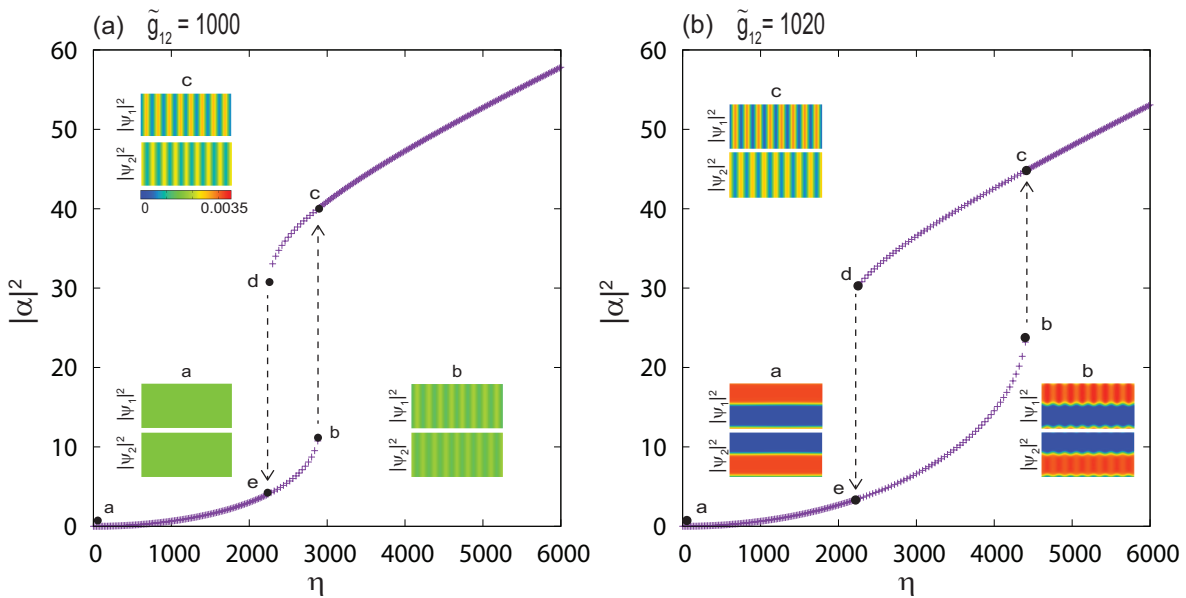} 
\caption{
   Steady-state intracavity photon number $|\alpha|^2$ as a function of laser pump intensity 
   $\eta$, where $\eta$ is increased from $\eta = 0$ to 6000 and 
   then decreased to 0 in the imaginary-time evolution. The density profiles of the 
   two components at points a, b, and c are shown in the insets, 
   where the size in the $y$ direction is reduced to 1/2.
   The parameters are $\tilde{g} = 1000$, 
   $ U_{1}=0.25 $, 
   $ U_{2}=0.125 $, $ \delta_{c}=1200 $, 
   $ \kappa=400 $, and $ N = 16 \tilde{N} = 1.2 \times 10^5 $. 
   (a) $ \tilde{g}{_{12}}=1000 $ and (b) $ \tilde{g}{_{12}}=1020 $.
\label{fig:fig1}}
\end{figure*}
First, in Fig.~\ref{fig:fig1}(a), we show the case of $ \tilde{g} = \tilde{g}_{12} $, for which the two components are miscible when $\eta = 0$.
We prepare the ground state for $\eta = 0$, and $\eta$ is gradually increased with a step of $\Delta \eta = 40$. For each value of $\eta$, imaginary-time propagation is performed for a long enough time ($\Delta T = 100$) that the system follows the steady state for each $\eta$. This process mimics experiments in which the pump strength is changed much more slowly than the relaxation time of the system. After $\eta$ reaches 6000, $\eta$ is decreased with a step of $\Delta\eta = -40$ to $\eta = 0$ to study the hysteresis. In Fig.~\ref{fig:fig1}(a), we find bistable behavior; the system jumps from the lower to upper branches at $\eta \simeq 2920$ when $\eta $ is increased, and the system jumps back to the lower branch at $\eta \simeq 2220$ when $\eta$ is decreased. The two components are mixed in the lower branch, while the alternate stripe pattern is formed in the upper branch, as shown in the insets in Fig.~\ref{fig:fig1}(a). 

In Fig.~\ref{fig:fig1}(b), we show the result for $\tilde{g}_{12}=1020 $ and $\tilde{g}=1000 $, which satisfy the immiscibility condition for the homogeneous system. When $\eta = 0$ (point a in Fig.~\ref{fig:fig1}(b)), the two components are spatially separated along the $y$ axis due to the immiscible condition. When $\eta$ is increased to 4400, the density profile of each component becomes modulated by the cavity field (inset b). At this value of $\eta$, the system jumps from point b to point c, where the density profiles of the two components suddenly change to the alternate stripe pattern. The system then jumps back to the separated state, when $\eta$ is decreased, exhibiting hysteresis. 
\begin{figure*}[tb]
\includegraphics[scale=0.45]{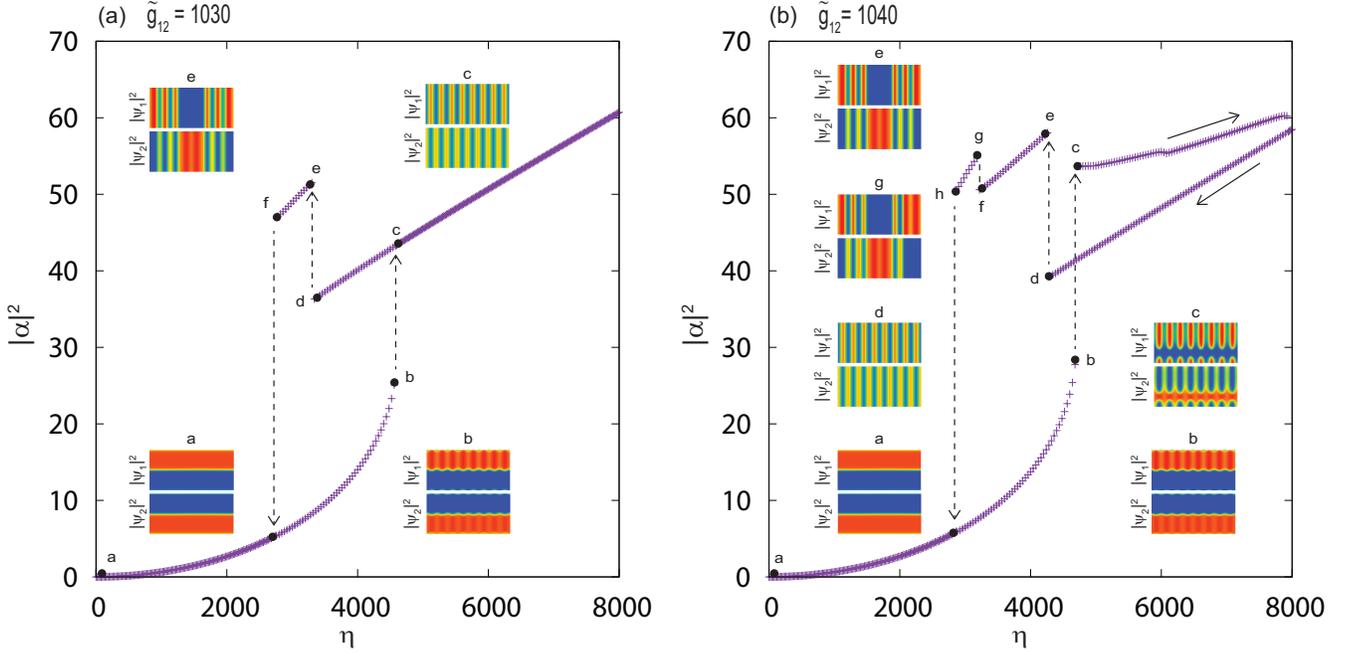}
\caption{Steady-state intracavity photon number $|\alpha|^2$ as a function of laser pump intensity $\eta$ for (a) $\tilde{g}_{12} = 1030$ and (b) $\tilde{g}_{12} = 1040$. The density profile at each point is shown in the inset. Other parameters are the same as those in Fig.~\ref{fig:fig1}.
\label{fig:fig3}}
\end{figure*}

We note that the bistable region is significantly increased from Fig.~\ref{fig:fig1}(a) to Fig.~\ref{fig:fig1}(b), whereas the parameters in 
Fig.~\ref{fig:fig1}(b) are the same as those in Fig.~\ref{fig:fig1}(a) except for a small change in the intercomponent interaction coefficient $\tilde{g}_{12}$. This is due to the miscibility-immiscibility transition, which alters the global structure of the atomic density distribution, i.e., the density distributions in the lower branches in Figs.~\ref{fig:fig1}(a) and~\ref{fig:fig1}(b) are quite different. The results in Figs.~\ref{fig:fig1}(a) and~\ref{fig:fig1}(b) thus demonstrate that the bistability curve is sensitive to $\tilde{g}_{12}$, which can be used to control the optical bistability, since $\tilde{g}_{12}$ can be controlled using the Feshbach resonance.

In Figs.~\ref{fig:fig3}(a) and~\ref{fig:fig3}(b), $\tilde{g}_{12}$ is further increased, and we observe the transition from bistability to multistability. In Fig.~\ref{fig:fig3}(a), for $\tilde{g}_{12} = 1030$, we find a new branch around $\eta \simeq 3000$, when $\eta$ is decreased. In this branch, the stripe phase and separated phase coexist, as shown in the inset e in Fig.~\ref{fig:fig3}(a), where the central region and outer region are occupied by components 2 and 1, respectively, and the stripe regions lie in between. This coexistence phase appears because it has lower effective energy than the overall stripe phase (inset c), which will be shown in Sec.~\ref{sec:fifth} using the variational method. 

Figure~\ref{fig:fig3}(b) shows the case of $\tilde{g}_{12}=1040$, where one can see five stable branches.
Initially, in the lower stable branch, the two components are strongly separated, as shown in the insets a and b. At $\eta \simeq 4680$, the system jumps from point b to point c, where the two regions coexist along the y axis: the stripe region and the region occupied only by component 2. This coexistence phase continues upto $\eta \simeq 8000$, at which the region occupied by component 2 disappears and the whole space becomes the stripe state. When $\eta$ is decreased, the stripe state continues to point d, and the transition occurs from point d to point e, which is similar to the transition in Fig.~\ref{fig:fig3}(a) (d to e).
After that, further transition occurs from point f to point g, where only the number of stripes changes. Such transition is ascribed to the finite size of the system, and will become continuous in an infinite system.
\section{Variational approach}
\label{sec:fourth}
In the previous section, we numerically showed that various multistable phases appear when the intercomponent repulsive interaction is increased. In order to understand this result in more detail, we performed variational analysis. The effective energy of the system can be written as \cite{zhou2009cavity}
\begin{equation}
\begin{aligned}
E_{\rm eff} = {} & \int d\bm{r}\Big(\vert\nabla\Psi_{1}\vert^{2}+\vert\nabla\Psi_{2}\vert^{2}+\dfrac{g_{11}N}{2}\vert\Psi_{1}\vert^{4}\\
& +\dfrac{g_{22}N}{2}\vert\Psi_{2}\vert^{4}+g_{12}N\vert\Psi_{1}\vert^{2}\vert\Psi_{2}\vert^{2}\Big)\\
& +\frac{\eta^{2}}{\kappa}\tan^{-1}\Bigg\{{\frac{1}\kappa}\Big[\delta_{c}+\frac{U_{1}N}{2}\int d\bm{r}\vert\Psi_{1}\vert^{2}\cos(2kx)\\
& +\frac{U_{2}N}{2}\int d\bm{r}\vert\Psi_{2}\vert^{2}\cos(2kx)\Big]\Bigg\}.
\end{aligned} 
\label{eq:fivea}
\end{equation}
This effective energy is defined in such a way that its functional derivative $\delta E_{\rm eff} / \delta\Psi^*(\bm{r})$ gives the right-hand sides of the GP equations~(\ref{eq:one}) and~(\ref{eq:two}) with Eq.~(\ref{eq:four}), and therefore minimization of $E_{\rm eff}$ globally or locally under the constraint of  the normalization of the wave functions gives the steady states of the GP equations.

From the density distributions shown in the insets in Figs.~\ref{fig:fig1} and \ref{fig:fig3}, we find that the steady states consist of different spatial regions (region occupied only by each component separately, alternate stripe region), and therefore we divide the integrals in Eq. (5) into $\sum_\beta I_\beta + \sum_{\beta \beta'} I_{\beta \beta'}$, where $\beta = a, b, \cdots$ is the index of the regions. In this expression, $I_\beta$ denotes the bulk part of the region $\beta$ and $I_{\beta \beta'}$ denotes the interface part between the regions $\beta$ and $\beta'$. Since the interface part is difficult to evaluate, we consider a sufficiently large system, in which the bulk parts are dominant and the interface parts can be neglected. The effective energy is then approximated as
\begin{equation}
\begin{aligned}
 E_{\rm eff} ={} & \sum_{\beta}\int_\beta d\bm{r}\Big( \vert\nabla\Psi_{1\beta}\vert^{2}+\vert\nabla\Psi_{2\beta}\vert^{2}\\
 &+\dfrac{g_{11}N}{2}\vert\Psi_{1\beta}\vert^{4}+\dfrac{g_{22}N}{2}\vert\Psi_{2\beta}\vert^{4}\\
 &+g_{12}N\vert\Psi_{1\beta}\vert^{2}\vert\Psi_{2\beta}\vert^{2}\Big)+\frac{\eta^{2}}{\kappa}\tan^{-1}\Bigg\{{\frac{1}\kappa}\Big[{\delta_{c}}\\
 &+\frac{U_{1}N}{2}\sum_{\beta}\int_\beta d\bm{r}\vert\Psi_{1\beta}\vert^{2}\cos(2kx)\\
 &+\frac{U_{2}N}{2}\sum_{\beta}\int_\beta d\bm{r}\vert\Psi_{2\beta}\vert^{2}\cos(2kx)\Big]\Bigg\},
\end{aligned} 
\label{eq:five}
\end{equation}
where $\Psi_{j\beta}$ represents the wave function of component $j$ in the region $\beta$ and the integrals $\int_\beta d\bm{r}$ are taken only in each region $\beta$.

Next, we introduce the two-mode approximation.
The ground state of the condensate in the absence of external pumping is homogeneous, with a zero-momentum state in each region $\beta$. The effect of the intracavity field is to diffract this ground state from the zero-momentum state into a superposition $\frac{1}{\sqrt{2}}(\ket{p=+2\hbar k}+\ket{p=-2\hbar k})$ of the momentum states. In a weak interaction and weak pump case, the state of the condensate is limited to these two modes.
We therefore approximate the wave function as
\begin{equation}
\Psi_{j\beta}(\bm{r})=\sqrt{\frac{N_{j\beta}}{V_{\beta}}}\Big[a_{j\beta}+\sqrt{2}b_{j\beta}\cos(2kx)\Big], 
\label{eq:fivea}
\end{equation} 
where $N_{j\beta}$ is the number of component-$j$ atoms in the region $\beta$, $V_\beta$ is the volume of the region $\beta$, and $a_{j\beta}$ and $b_{j\beta}$ are real numbers satisfying $a_{j\beta}^2 + b_{j\beta}^2 = 1$. We also define the atom number ratio $ n_{j\beta} =  N_{j\beta} / N_j$ and the volume ratio $v_\beta = V_\beta / V$, where $V$ is the total volume. These ratios must satisfy $\sum_\beta n_{j\beta} = \sum_\beta v_\beta = 1$. Substituting Eq.~(\ref{eq:fivea}) into Eq.~(\ref{eq:five}), we obtain the variational energy as
\begin{equation}
\begin{aligned}
\frac{E_{\rm eff}}{N} ={} & 4\sum_{j=1,2}\sum_{\beta} n_{j\beta}b_{j\beta}^{2}\\
&+\sum_{j=1,2}\sum_{\beta}\frac{g_{jj} N}{V}\frac{n_{j\beta}^{2}}{v_\beta}\Big(1+4a_{j\beta}^{2}b_{j\beta}^{2}+\frac{1}{2}b_{j\beta}^{4}\Big)\\
&+\frac{g_{12} N}{V}\sum_{\beta}\frac{n_{1\beta}n_{2\beta}}{v_\beta}\Big(1+4a_{1\beta}b_{1\beta}a_{2\beta}b_{2\beta}\\
&+\frac{1}{2}b_{1\beta}^{2}b_{2\beta}^{2}\Big)+\dfrac{\eta^{2}}{\kappa N}\tan^{-1}\Bigg[\frac{1}{\kappa}\Big(\delta_{c}\\
&+\sum_{j=1,2}\sum_{\beta}\frac{1}{\sqrt{2}}NU_{j}n_{j\beta}a_{j\beta}b_{j\beta}\Big)\Bigg],
\end{aligned} 
\label{eq:six}
\end{equation}
\begin{table*}
\caption{\label{tab:table1}Variational parameters for the five phases.}
\begin{ruledtabular}
\begin{tabular}{cccccccccc}
 Phase&$ n_{1a} $&$ n_{2a} $ & $ n_{1b} $&$ n_{2b} $&$ n_{1c} $&$ n_{2c} $&$ v_a $&$ v_b $&$ v_c $\\ \hline
 $1$ & $1$ & $0$ & $0$ & $1$ & $0$ & $0$ & variable & $1-v_a$ & $0$ \\
 $2$ & $1$ & $1$ & $0$ & $0$ & $0$ & $0$ & $1$ & $0$ & $0$\\
 $3$ & $1$ & variable & $0$ & $1 - n_{2a}$ & $0$ & $0$ & variable & $1-v_a$ & $0$\\
 $4$ & variable & $1$ & $1 - n_{1a}$ & $0$ & $0$ & $0$ & variable & $1-v_a$ & $0$\\
 $5$ & variable &  variable & $0 $ & $1 - n_{2a}$ & $1 - n_{1a}$& $ 0 $ & variable & variable & $1-v_a-v_b$\\
\end{tabular}
\end{ruledtabular}
\end{table*}
where the variational parameters are $n_{j\beta}$, $v_\beta$, and $b_{j\beta} = \sqrt{1 - a_{j\beta}^2}$. To compare the variational results with those in Sec. III, we use the same parameters $g$, $U_j$, $\delta_c$, $\kappa$, and $N$. As in Sec. III, we define $\tilde{g}_{ij} = g_{ij} N / 4$ and we set $V = (8\pi)^2$. The thermodynamic limit $ N,V\rightarrow \infty $ can be taken with $ U_{j}N  $ and
 $\eta^{2}/N $ being kept constant. 
 \begin{figure}[tb]
\includegraphics[scale=0.45]{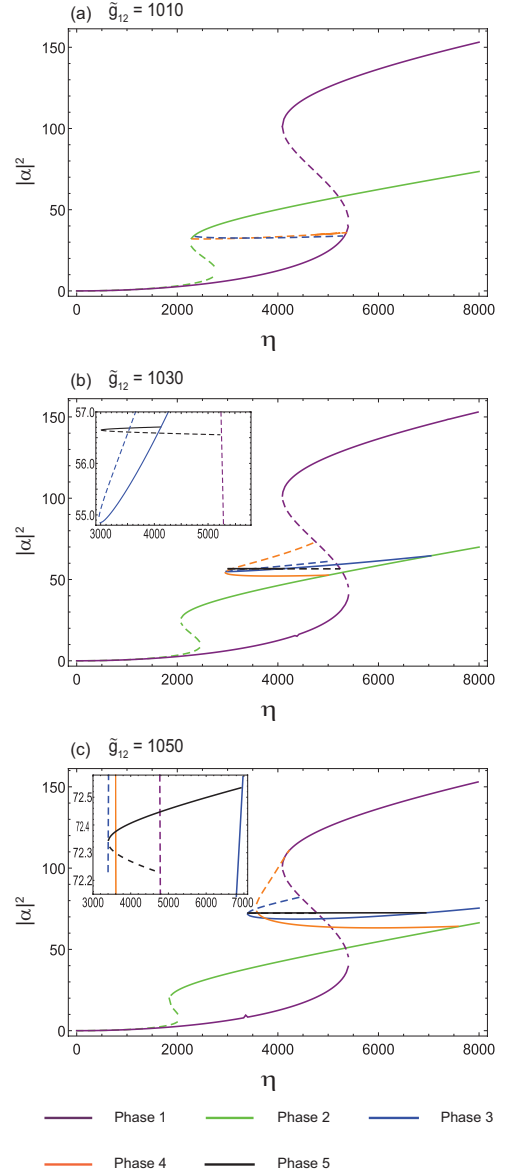} 
\caption{
Steady-state intracavity photon number $|\alpha|^2$ as a function of laser pump intensity $\eta$ for (a) $\tilde{g}_{12} = 1010$, (b) $\tilde{g}_{12}= 1030$, and (c) $\tilde{g}_{12} = 1050$. The solid and dashed lines indicate energetically stable and unstable states, respectively. The insets provide magnification. Other parameters are $ \tilde{g}_{11} = \tilde{g}_{22} = 1000 $, $ N=1.2 \times 10^5 $, $ U_{1}=0.25 $, $ U_{2}=0.125 $ , $ \delta_{c}=1200 $, and
   $ \kappa=400 $.
   }
\label{fig:fig9}
\end{figure}

To find the stationary conditions of the variational parameters, we first minimize Eq.~(\ref{eq:six}) by the Monte Carlo method, in which the initial values of variational parameters are set to random numbers, and they are changed in a random-walk-like manner in the parameter space to reach the minima. By this method, we obtain an initial guess of possible stationary phases, which are summarized in Table I. There are five phases as follows:\\
Phase 1: Two components are separated in two spatial regions $\beta = a$ and $b$, and therefore $n_{1a} = n_{2b} = 1$ and $n_{1b} = n_{2a} = 0$.\\
Phase 2: The alternate stripe pattern is formed in the whole space, and there is only a single region $\beta = a$.\\
Phase 3: There are two regions $\beta = a$ and $b$. Region $a$ is the alternate stripe state and region $b$ is occupied only by component 2, and hence $n_{1a} = 1$ and $n_{1b} = 0$.\\
Phase 4: Similar to phase 3, but the region $b$ is occupied only by component 1.\\
Phase 5: We have three regions $\beta = a$, $b$, and $c$. Region $a$ is the alternate stripe state, region $b$ is occupied only by component 2, and region $c$ is occupied only by component 1.\\
By the Monte Carlo search, no other phases are found, and three spatial regions are sufficient.
From the initial guess of the stationary states, we can obtain the complete stationary curves by an iteration method such as the Newton-Raphson method. The energetically stable states can be obtained by both Monte Carlo and iteration methods, while unstable states are obtained only by the iteration method. These results are shown in Fig.~\ref{fig:fig9}.

\begin{figure}[tb]
\includegraphics[scale=0.5]{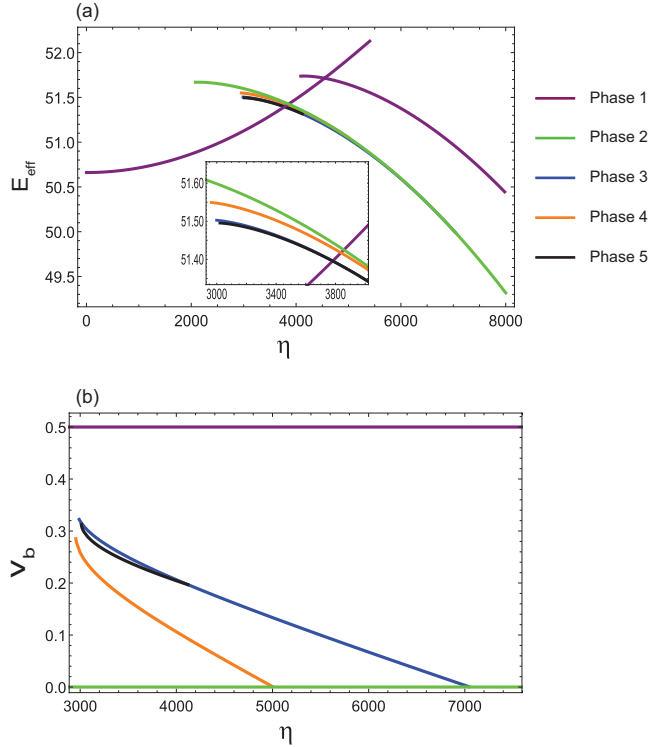} 
\caption{ 
(a) Effective energy $E_{\rm eff}$ as a function of laser pump intensity $\eta$.
(b) Volume ratio of region b, $v_b$, as a function of laser pump intensity $\eta$.
The intercomponent interaction strength is $\tilde{g}_{12} = 1030  $. Other parameters are the same as those in Fig.~\ref{fig:fig9}. Curves of stable states are only shown.
   }
\label{fig:fig10}
\end{figure}

Figure~\ref{fig:fig9}(a) shows the case of $\tilde{g}_{12} = 1010$. Unlike the single bistable curve in a usual BEC-cavity system, one can see two bistable curves in Fig. 4(a), which arise from the separated phase and alternate stripe phase, respectively. For small $\eta$, the only stable branch is the separated phase (phase 1). Increasing $\eta$, the lower stable branch of phase 1 disappears at $\eta \simeq 5400$, above which the state can jump to either upper branch of phase 1 or phase 2. Since the energy of phase 2 is lower than that of phase 1 around this value of $\eta$ (see Fig.~\ref{fig:fig10}(a)), phase 2 is chosen at this jump, i.e., the stripe state is realized. When $\eta$ is decreased to 
$\eta\simeq 2280$, the upper stable branch of phase 2 disappears and the state jumps back to phase 1. Thus, Fig.~\ref{fig:fig9}(a) can explain the behavior of the numerical result in Fig.~\ref{fig:fig1}(b), and reveals that the bistable behavior in Fig.~\ref{fig:fig1}(b) arises from double S-shaped curves, not from a single S-shaped curve as usual. 

We increase $\tilde{g}_{12}$ to 1030 and 1050, which are shown in Figs.~\ref{fig:fig9}(b) and~\ref{fig:fig9}(c), respectively. Phases 3, 4, and 5 additionally appear as stable phases, which leads to multistability. One can see that the stable parts of phases 3 and 4 connect to phase 2, and the stable part of phase 5 connects to phase 3, exhibiting complicated structures. In the numerical result in Fig. 3(a), the coexistence state appeared (inset e) in a narrow range of $\eta$, which corresponds to phase 5. This is consistent with Fig.~\ref{fig:fig9}(b), where the stable part of phase 5 only appears in a narrow range of $\eta$. The emergence of phase 5 can also be understood from Fig.~\ref{fig:fig10}(a), which shows that phase 5 has lower effective energy than phase 2.

The behavior in Fig.~\ref{fig:fig3}(b) is also consistent with the variational result in Figs.~\ref{fig:fig9}(b) and~\ref{fig:fig9}(c). In Fig.~\ref{fig:fig3}(b), the jump occurred from point b to point c, which corresponds to the jump from phase 1 to phase 3. This is consistent with Fig.~\ref{fig:fig9}, i.e., the stable range of phase 3 is increased by an increase of $\tilde{g}_{12}$, which enables the jump from the lower edge of phase 1 to phase 3. As $\eta$ is increased, phase 3 merges into phase 2 in Figs.~\ref{fig:fig9}(b) and~\ref{fig:fig9}(c), which also agrees with the behavior in Fig.~\ref{fig:fig3}(b). The volume ratio of region b decreases with increasing $\eta$ as shown in Fig.~\ref{fig:fig10}(b).

\begin{figure}[tb]
\includegraphics[scale=0.5]{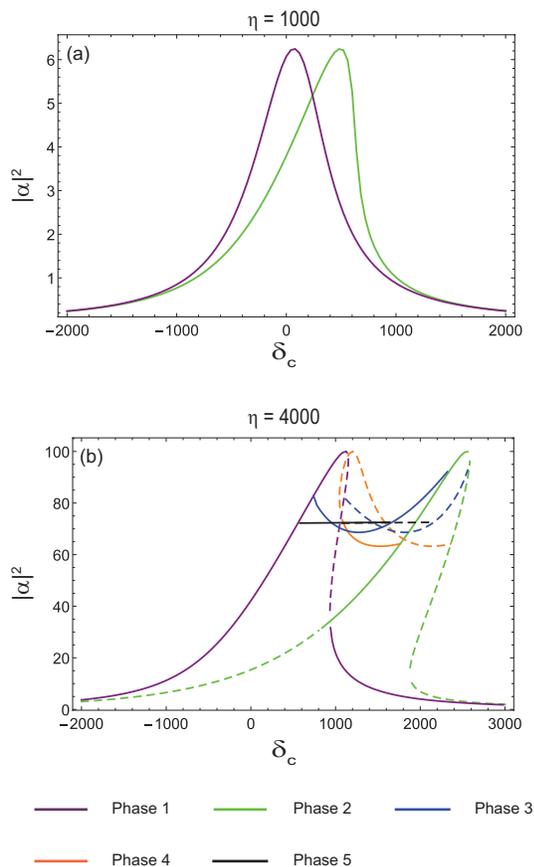} 
\caption{
 Steady-state intracavity photon number $|\alpha|^2$ as a function of effective detuning $\delta_{c}$ for (a) $\eta=1000$ and (b) $\eta=4000$. The value of intercomponent interaction is $ \tilde{g}_{12} = 1050 $. Other parameters are the same as those in Fig.~\ref{fig:fig9}.
   }
\label{fig:fig12}
\end{figure}

Figure~\ref{fig:fig12} shows the $\delta_c$ dependence of the photon number with $\eta$ fixed.
When the pump intensity is $\eta=1000$, the stationary curves of phases $1$ and $2$ are nearly Lorentzian, as shown in Fig.~\ref{fig:fig12}(a). When the pump increases, the atoms collectively act as a dispersive medium shifting the cavity resonance, and the system exhibits multistable behavior above a critical value, as shown in Fig.~\ref{fig:fig12}(b) for $\eta=4000$, where the Lorentzian-like curves shift rightward. Furthermore, the stationary curves of phases 3, 4, and 5 also emerge, which connect to those of phases 1 and 2, forming complicated structures.

\section{\label{sec:fifth}Trapped system}

We have so far considered a two-dimensional system without an external potential. Here we examine the multistability for a more realistic three-dimensional system confined in a harmonic potential. We consider a two-component BEC with different hyperfine states of $^{87}$Rb in a trap potential $ V_1(\bm{r}) = V_2(\bm{r}) = m (\omega_x^2 x^2 + \omega_y^2 y^2 + \omega_z^2 z^2) / 2$, where the trap frequencies are $\omega_x = 2\pi \times 800$Hz, $\omega_y = 2\pi \times 8000$Hz, and $\omega_z = 2\pi \times 800$Hz. We normalize the length and time by $k^{-1}$ and $(\hbar k^2 / 2m)^{-1}$ with $2\pi / k = 780$ nm. We solve the imaginary-time evolution of the GP equation with increasing and decreasing $\eta$ by the step $\Delta\eta = \pm 80$ and obtain the stationary state for each $\eta$, as in Sec. III.
\begin{figure}[tb]
\includegraphics[scale=0.5]{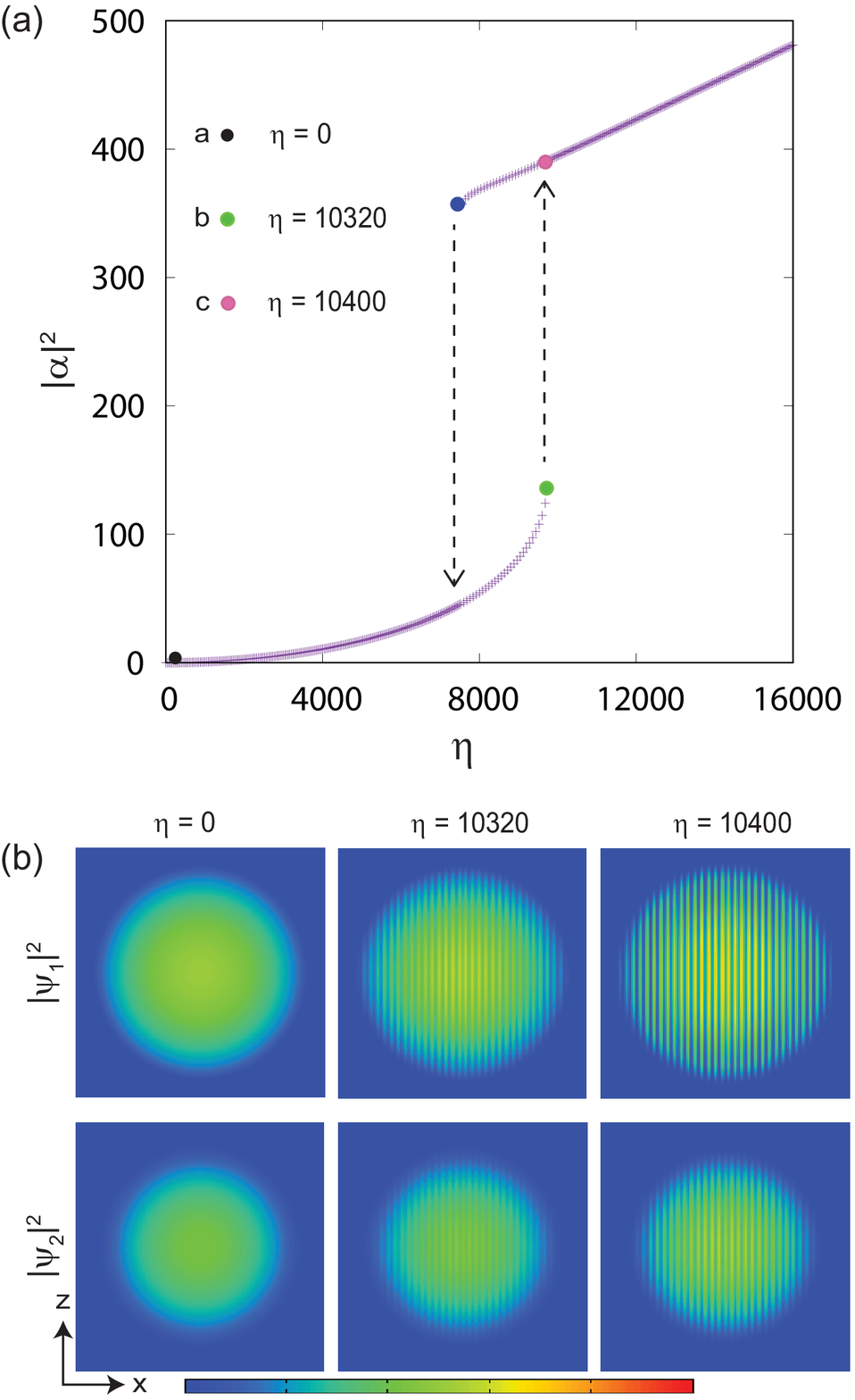}
\caption{(a) Steady-state intracavity photon number $|\alpha|^2$ as a function of laser pump intensity $\eta$ for $g_{12} = 0.9 g$. (b) Density profiles of two components on the $y = 0$ plane. The size of each panel is $115.2\mu{\rm m}$ $\times$ $115.2\mu{\rm m}$ and the color bar ranges from 0 to $5 \times10^{-5}$ in units of $N k^3$. Each colored point in (a) corresponds to the density profile in (b) for each value of $\eta$. 
The parameters are $a_{11}=100 a_{0}$, with $a_0$ being the Bohr radius, $U_{1}=0.025 $, $U_{2}=0.0125 $ , $\delta_{c}=1200 $, and $ N=6\times10^{5} $.}
\label{fig:fig13}
\end{figure}

Let us first consider the case of $ g \leq g_{12} $. In Fig.~\ref{fig:fig13}, we show the stationary states as a function of $\eta$. For $\eta = 0$, the two components are mixed. As $\eta$ is increased, the system follows the lower branch, and jumps to the upper branch at $\eta \simeq 10400$, where the alternate stripe pattern emerges. When we decrease $\eta$, the system jumps from the upper to lower branches at $\eta \simeq 7880$, resulting in the bistable region. Such bistable behavior is similar to that in the ideal system in Fig.~\ref{fig:fig1}(a).

\begin{figure}[tb]
\includegraphics[scale=0.38]{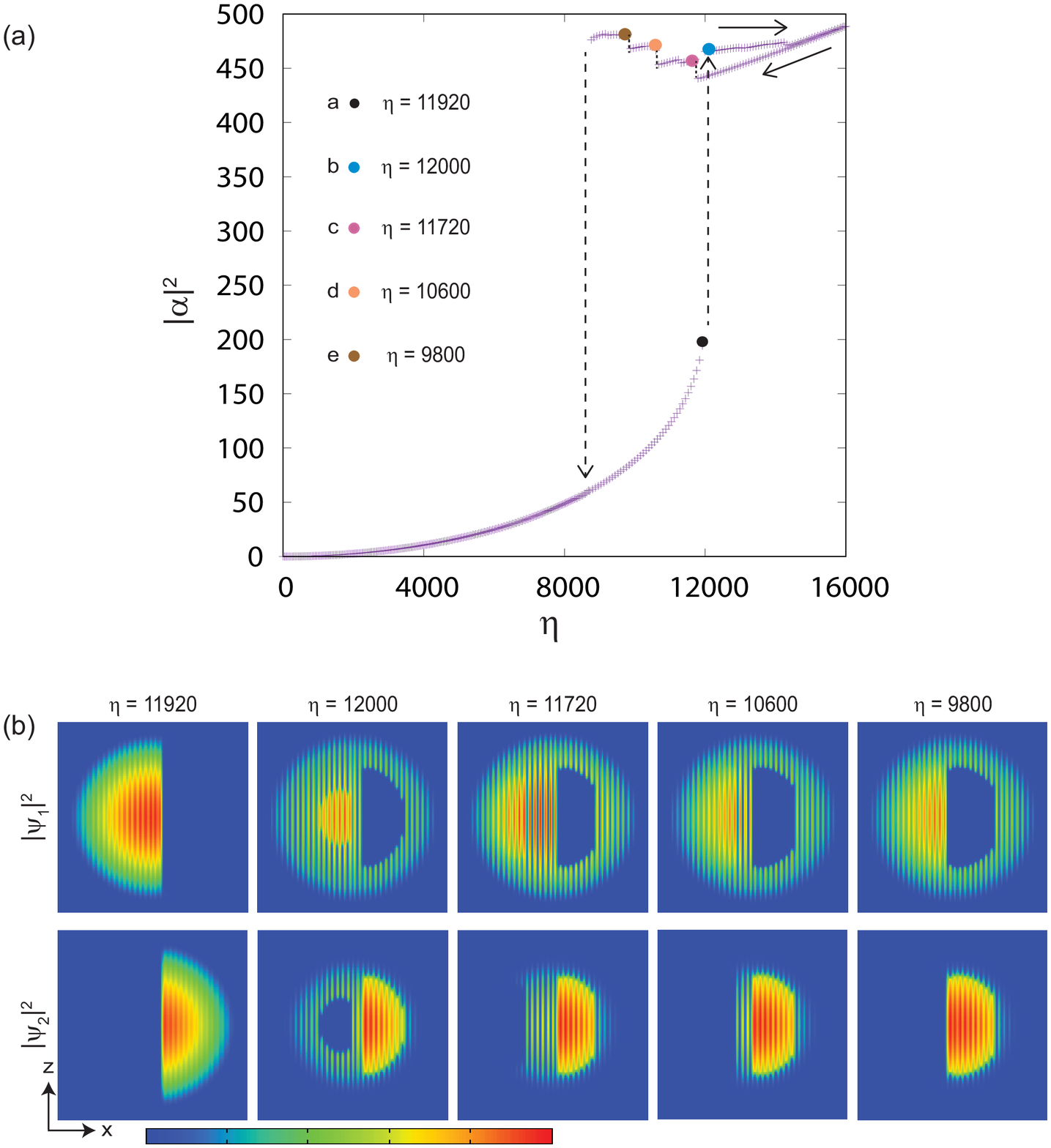}
\caption{(a) Steady-state intracavity photon number $|\alpha|^2$ as a function of laser pump intensity $\eta$ for $g_{12} > g$. (b) Density profiles of two components on the $y = 0$ plane. The size of each panel is $115.2\mu{\rm m}$ $\times$ $115.2\mu{\rm m}$ and the color bar ranges from 0 to $5 \times10^{-5}$ in units of $N k^3$. Each colored point in (a) corresponds to the density profile in (b) for each value of $\eta$. The intercomponent scattering length is $ a_{12}=110 a_{0}$, where $a_0$ is the Bohr radius. All other parameters are the same as those in Fig.~\ref{fig:fig13}.}
\label{fig:fig14}
\end{figure}

Figure~\ref{fig:fig14} shows the immiscible case, $ g_{12}>g$. In this case, the lower branch starting from $ \eta=0 $ is the separated state. At $\eta \simeq 12000$, the system jumps to the upper branch, and the condensate sharply changes to a state in which the alternate stripe region and the regions occupied only by either component separately coexist. In this branch, the inner and outer regions are partially occupied by component 2 and 1. This phase is maintained upto $\eta=16000$. When $\eta$ is decreased from 16000, the system follows a different branch in a manner similar to Fig.~\ref{fig:fig3}(b). At $\eta \simeq 11720$, another jump occurs, and the stripe region appears around the center. Subsequently, two jumps occur in which the central stripe region decreases, resulting in the totally separated state at $\eta \simeq 9800$. By further decreasing $\eta$, the system jumps back to the lower branch, which is separated in the $x$ direction.

Thus, in the trapped three-dimensional system as well, bistability and multistability emerge, which is significantly altered by the miscibility of the two-component BEC. 

\section{\label{sec:sixth}conclusions}
We have investigated a two-component BEC coupled to a single-mode optical cavity. 
We showed that this coupled BEC-cavity system exhibits a variety of density structures and optical multistability, which is sensitive to the intercomponent interaction strength. This sensitivity arises from the miscible-immiscible transition of the two-component BEC, by which the global structure of the density distribution changes, altering the effective BEC-cavity coupling significantly. We found a variety of stable phases, such as the separated phase, alternate stripe phase, and their coexistence, which exhibit multistability. Using a variational method, we classified these phases, and explained the results of the numerical simulations.

In contrast to the single component BEC-cavity system, such a two-component BEC-cavity system allows one to control the optical bistability width by changing the condition of miscibility. Experimentally, the miscibility transition can be controlled by adjusting the $s$-wave scattering length through a Feshbach resonance.
Thus, the present system may provide a candidate for a controlled optical switch using the miscibility transition. The hydrodynamic properties of quantum fluids, such as their miscibility, will yield other interesting phenomena in the BEC-cavity system, which merit further study.
\begin{acknowledgments}
This work was supported by JSPS KAKENHI Grant Number JP20K03804.
\end{acknowledgments}
\appendix

%

\end{document}